\documentclass{PoS}

\usepackage{amsmath}
\usepackage{graphicx}

\newcommand{\nc}{\newcommand}
\nc{\tr}{{\rm{Tr}}}
\nc{\retr}{{\rm{Re\, Tr}}}
\nc{\dg}{\dagger}
\nc{\hm}{\hat{\mu}}
\nc{\hn}{\hat{\nu}}
\nc{\eq}[1]{~(\ref{#1})}
\nc{\Fmunu}{F_{\mu\nu}}
\nc{\Dmu}{{\mathcal D}_\mu}
\nc{\Dnu}{{\mathcal D}_\nu}
\nc{\fundDmu}{D_\mu}
\nc{\fundDnu}{D_\nu}
\nc{\scD}{{\mathcal D}}

\nc{\FSq}{\Fmunu^2(x)}
\nc{\DmuFSq}{(\Dmu\Fmunu(x))^2}
\nc{\DnuFSq}{(\Dnu\Fmunu(x))^2}
\nc{\DmuSqFSq}{(\Dmu^2\Fmunu(x))^2}
\nc{\DnuSqFSq}{(\Dnu^2\Fmunu(x))^2}
\nc{\DmuSqFDnuSqF}{\Dmu^2\Fmunu(x)\Dnu^2\Fmunu(x)}
\nc{\DmuDnuFSq}{(\Dmu\Dnu\Fmunu(x))^2}
\nc{\FSqSq}{\Fmunu^4(x)}

\title{Aspects of QCD Vacuum Structure}

\ShortTitle{Aspects of QCD Vacuum Structure}

\author{Peter J. Moran and Derek B. Leinweber\thanks{
We thank the Australian Partnership for Advanced Computing (APAC) and
the South Australian Partnership for Advanced Computing (SAPAC) for
generous grants of supercomputer time which have enabled this project.
This work is supported by the Australian Research Council.
}\\
        Special Research Centre for the Subatomic Structure of Matter
        and Department of Physics, University of Adelaide, SA 5005,
        Australia \\
        E-mail: \email{peter.moran@adelaide.edu.au}, \\
	\hspace*{1.18cm}\email{dleinweb@physics.adelaide.edu.au}
        }

\abstract{ 

The impact of dynamical fermions on the vacuum structure of QCD is
explored.  Of particular interest is the topological charge
correlator, $\langle q(x) q(0) \rangle$, where negative values at
small $x$ reveal a sign-alternating layered structure to the
topological-charge density of the QCD vacuum.  We consider large $28^3
\times 96$ lattices from the MILC collaboration, and develop a new
gluonic definition of the topological charge density, founded on a new
over-improved stout-link smearing algorithm.  The algorithm reproduces
established results from the overlap formalism and is designed to
preserve instantons.  We examine the extent to which instanton-like
objects are found on the lattice.  Finally, we investigate the effects
of dynamical sea-quark degrees of freedom on topology and find that
the magnitudes of the negative dip in the $\langle q(x)q(0) \rangle$
correlator and the positive $\langle q(0)^2 \rangle$ contact term are
both increased with the introduction of dynamical fermion degrees of
freedom.  This is in accord with expectations based on charge
renormalization and the vanishing of the topological susceptibility in
the chiral limit.

}

\FullConference{The XXV International Symposium on Lattice Field Theory\\
		 July 30-4 August 2007\\
		 Regensburg, Germany}

\begin{document}

\section{Introduction}

Understanding the topological structure of the QCD vacuum remains a
central focus of modern Lattice QCD studies. For computational
reasons, most previous studies have focused on gauge fields generated
using the quenched approximation. In the following proceedings, we
present a quantitative comparison of vacuum structure for quenched and
dynamical-fermion gauge fields.

The introduction of fermion loops into the QCD action renormalizes the
coupling and demands smaller values for $\beta$ in obtaining the same
lattice spacing, $a$.  Smaller $\beta$ values will admit rougher gauge
fields such that we expect to see a higher density of non-trivial
topological excitations, particularly for lighter sea-quark masses.
For example, we anticipate larger values for the mean-square
topological charge density $\langle q^2(x) \rangle_x$.  This combined
with the vanishing of the topological susceptibility in the chiral
limit leads to our prediction that the negative dip in the topological
charge density correlator, $\langle q(x)q(0) \rangle$, will be
enhanced in full QCD with light dynamical-fermions.  To the best of
our knowledge, this is the first study of the $\langle q(x)q(0)
\rangle$ correlator in full QCD.

In order to study these differences in vacuum structure on the very
large MILC lattices, we commence with the development of a new gluonic
definition of the topological charge density, founded on a new form of
over-improved~\cite{GarciaPerez:1993ki} stout-link
smearing~\cite{Morningstar:2003gk} algorithm, designed to stabilize
instantons.  We then examine the extent to which instanton-like
objects are found on the lattice.  Finally, we investigate the effects
of dynamical sea-quark degrees of freedom on topology and find that
the magnitudes of the negative dip in the $\langle q(x)q(0) \rangle$
correlator and the positive $\langle q(0)^2 \rangle$ contact term are
both increased with the introduction of dynamical fermion degrees of
freedom.  The effect is significant and is easily observed in the
visualizations of the topological charge density provided at
the close of these proceedings.

\section{Over-Improved Stout-Link Smearing}

The removal of short-distance UV fluctuations is an important aspect
of defining the topological charge of a rough gauge-field
configuration.  For gluonic topological charge operators, one often
applies iterative smoothing algorithms which hold the risk of
destroying the very structures one hopes to reveal.
The corrosion of topological excitations in the QCD vacuum under
smoothing is due to the presence of discretization errors in the
approximation of the action. 
In the past there have been attempts
to remove these errors via the combination of different sized Wilson
loops in the calculation of the local action. When combining these
loops, one must carefully choose the coefficients of the different
shapes in order to cancel the leading order error terms, thereby
resulting in an \emph{improved}
action~\cite{Bonnet:2001rc,BilsonThompson:2002jk}.

Despite the improvements, improved actions can still spoil
instantons~\cite{Bilson-Thompson:2001ca}.  Consider, for example the
Symanzik $O(a^2)$ improved action, composed of the plaquette
($P_{\mu\nu}$) and rectangular ($R_{\mu\nu},\,R_{\nu\mu}$) Wilson
loops.
\begin{equation}
  S_{S} = \beta \sum_x \sum_{\mu<\nu} \bigg[
  \frac{5}{3} ( 1 - P_{\mu\nu}(x) ) 
  - \frac{1}{12} \big( ( 1 - R_{\mu\nu}(x) ) + ( 1 - R_{\nu\mu}(x) ) \big) 
  \bigg] \, ,
  \label{eqn:symanzikaction}
\end{equation}
We can Taylor expand the Symanzik action in orders of $a$ and
%
%
%
following Perez, \emph{et al.}~\cite{GarciaPerez:1993ki} substitute
the classical instanton solution~\cite{Belavin:1975fg}
\begin{equation}
  A_\mu(x) = \frac{x^2}{x^2 + \rho^2} \left( \frac{i}{g} \right)
             \partial_\mu (S) \, S^{-1} \, , \ \ \ \ \ \ 
  S \equiv \frac{x_4 \pm i \, \vec{x} \cdot \vec{\sigma} }{ \sqrt{x^2}} \, ,
  \label{eqn:instantonsoln}
\end{equation}
into the expanded action to find 
\begin{equation}
  S_{S}^{inst} = \frac{8 \pi^2}{g^2} \left[ 1 - \frac{17}{210}\left( \frac{a}{\rho} \right)^{\!\!4} \right] \, .
  \label{eqn:symanzikinst}
\end{equation}
The negativity of the $O(a^4)$ error means that this action will
destabilize instantons when used in an iterative scheme.  This occurs
because the smoothing algorithms are designed to remove action and
will do so by effectively reducing $\rho$ to obtain a lower action.
Eventually the instantons become sufficiently small that
discretization errors allow them to be removed from the lattice.

Perez, \emph{et al.}~\cite{GarciaPerez:1993ki} proposed that instead
of combining different loop combinations in order to suppress the
discretization errors, they could instead tune their coefficients such
that the errors became positive for a classical instanton. By
doing this, instantons should be stable under cooling. We extend
their work, using a plaquette plus rectangle action, in the interests
of locality, and modern stout-link smearing~\cite{Morningstar:2003gk}.

Taking the Symanzik action\eq{eqn:symanzikaction} and introducing
a new parameter $\epsilon$, such that $\epsilon = 1$ provides
the Wilson action and $\epsilon = 0$ provides the Symanzik-improved action,
implies the following form for the over-improved action
\begin{equation}
  S(\epsilon)  = \beta \sum_x \sum_{\mu<\nu} \bigg[
  \frac{5-2\epsilon}{3} ( 1 - P_{\mu\nu}(x) ) 
  - \frac{1-\epsilon}{12} \big( ( 1 - R_{\mu\nu}(x) ) + ( 1 - R_{\nu\mu}(x) ) \big) 
  \bigg] \, .
  \label{eqn:overimpaction}
\end{equation}
Taylor expanding this action for the classical instanton solution,
one finds that for $\epsilon < 0$ the leading order $a^2$ errors are
positive
\begin{equation}
  S^{inst}(\epsilon) = \frac{8 \pi^2}{g^2} \left[ 1 
    - \frac{\epsilon}{5} \left(\frac{a}{\rho}\right)^{\!\!2}
    + \frac{14\epsilon-17}{210} \left(\frac{a}{\rho}\right)^{\!\!4} \right] \, .
  \label{eqn:overimpinstaction}
\end{equation}
The question is now:
How negative should $\epsilon$ be? To answer this, we propose the
following method.

\begin{figure}[t]
  \begin{center}
    \includegraphics[angle=90,width=0.45\textwidth]{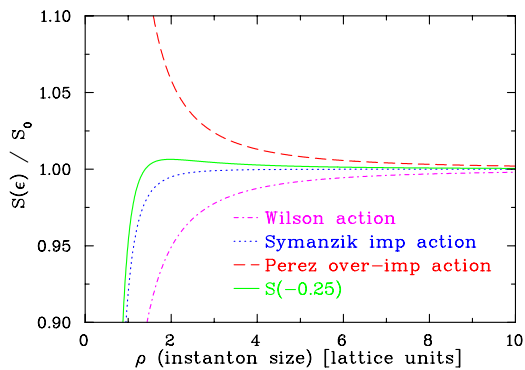}
  \end{center}
  \caption{$S(\epsilon)/S_0$ versus the instanton size, $\rho$, for
the Wilson action, Symanzik-improved action, Perez over-improved action,
and our over-improved action $S(-0.25)$. $S_0$ is the action
for a single instanton.  To preserve instantons the
ideal smoothing action would give a straight line at $S(\epsilon)/S_0 = 1$.
The slope of each curve dictates how an instanton in the gauge field
will evolve under smearing. 
Our action should be the most stable because it is mostly flat
and has a dislocation threshold, given by its maximum, at $\rho \sim 1.5$.}
  \label{fig:SvsRho}
\end{figure}

Given $S(\epsilon)$, select some value of $\epsilon$ and plot
$S(\epsilon)/S_0$ as a function of $\rho$. Ideally this will result in
a straight line at $S(\epsilon)/S_0 = 1$. What we actually observe is
illustrated in Fig.~\ref{fig:SvsRho}.  Note that it is the value of
the slope of the curve that is important when deciding how an
instanton will change under a given smoothing algorithm.  Varying
$\epsilon$ results in curves of varying slope.  We settled on a value
of $\epsilon = -0.25$ as providing a nice result.

\section{Vacuum Structure}

With the over-improved stout-link smearing procedure completely defined
we now proceed to perform a study of topological excitations in the QCD
vacuum. We also provide a few results obtained using a 3-loop
improved cooling algorithm~\cite{BilsonThompson:2002jk}.
We use one set of quenched gauge fields
and two sets of dynamical gauge fields 
in order to investigate the effect of dynamical sea quarks and varying
quark mass. The gauge fields were generated by the MILC
collaboration~\cite{Bernard:2001av,Aubin:2004wf}, and their details are summarized in
Table~\ref{table:latticedetails}.
\begin{table}[b]
  \caption{\label{table:gaugefields} Parameters of the gauge fields used for this investigation. Label denotes how we will refer to the respective lattices throughout this proceeding. For more information see~\cite{Bernard:2001av,Aubin:2004wf}.}

  \newcommand\T{\rule{0pt}{2.8ex}}
  \newcommand\B{\rule[-1.4ex]{0pt}{0pt}}

  \begin{center}
    \begin{tabular}{lccccr}
      \hline
      \hline
      label & size & $\beta$ & $a$ (fm) & $a m_{u,d}$ / $a m_s$ \\
      \hline
      Quenched & $28^3 \times 96$ & 8.40 & $0.086$ & $-$  \\
      Heavy & $28^3 \times 96$ & 7.11 & $0.086$ & $0.0124$ / $0.031$ \\
      Light & $28^3 \times 96$ & 7.09 & $0.086$ & $0.0062$ / $0.031$ \\
      \hline
    \end{tabular}
  \end{center}
  \label{table:latticedetails}
\end{table}

\subsection{Topological Charge Density Correlator}

\begin{figure}[t]
  \begin{center}
    \begin{tabular}{l r}
      \includegraphics[angle=90,width=0.45\textwidth]{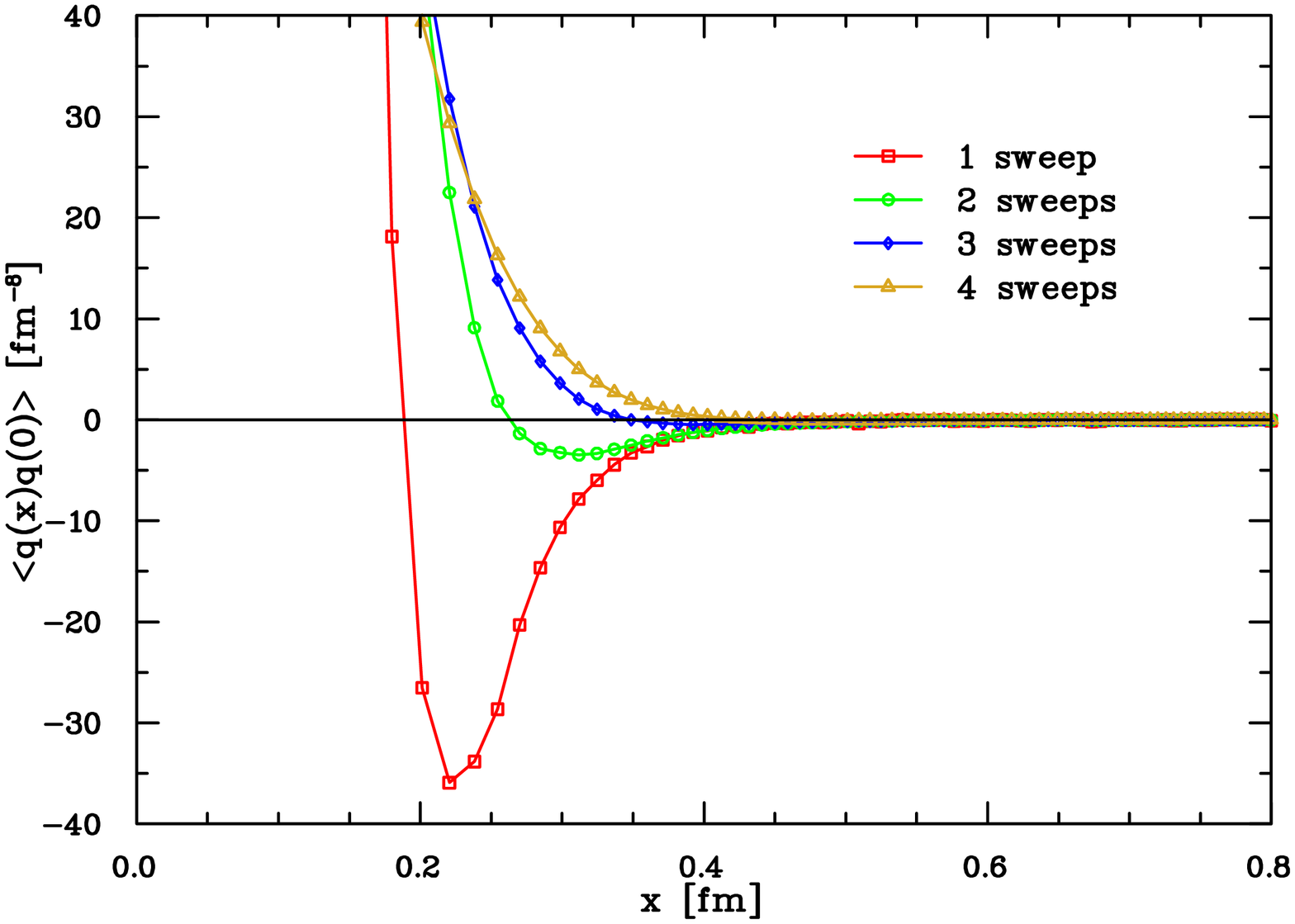}
      &
      \includegraphics[angle=90,width=0.45\textwidth]{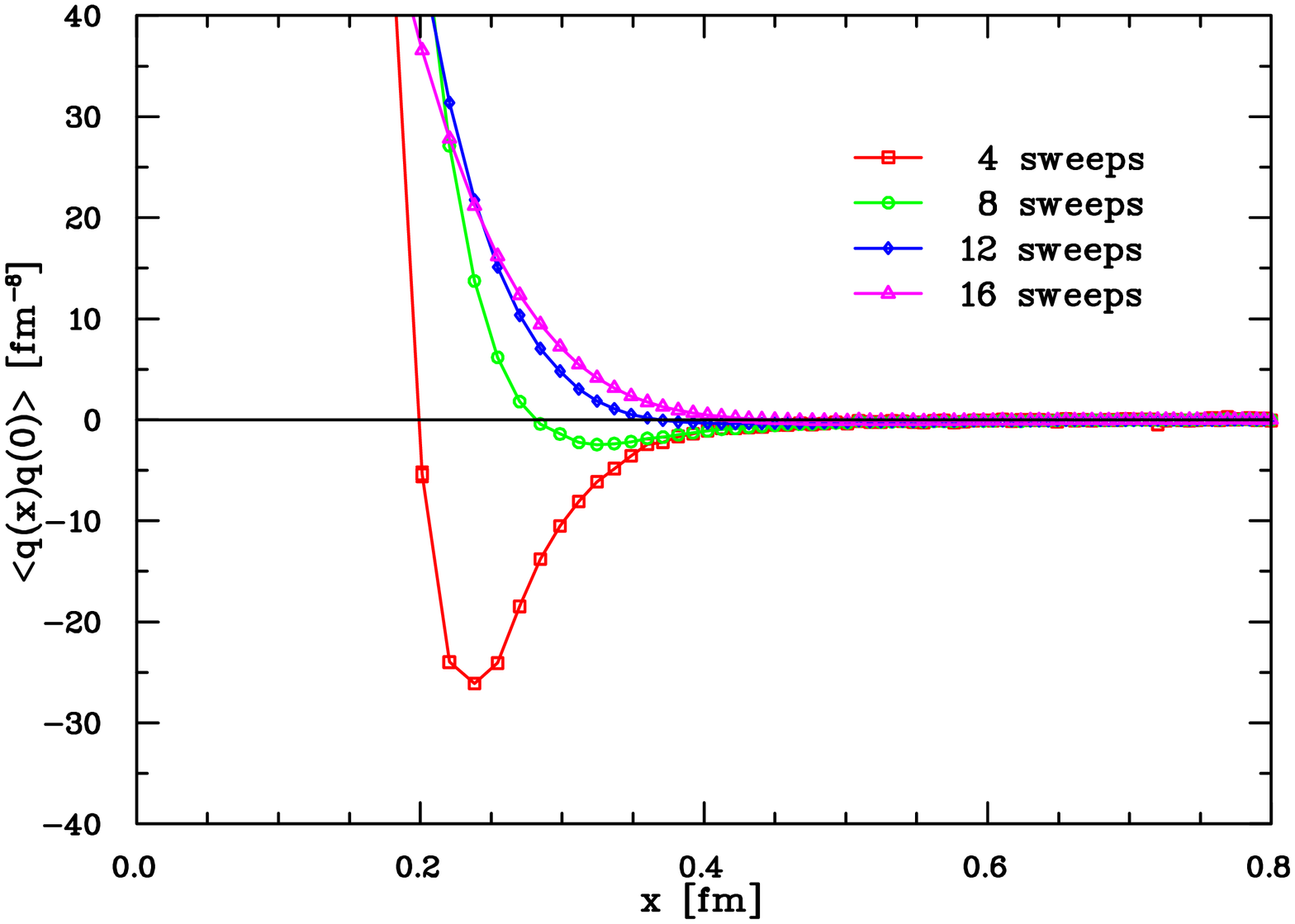}
    \end{tabular}
  \end{center}
  \caption{The topological charge density correlator $\langle q(x)q(0) \rangle$
as computed on the quenched gauge configurations for both 3-loop improved
cooling (left) and over-improved stout-link smearing (right).
We see that for a small number of sweeps it is
possible to obtain a negative $\langle q(x)q(0) \rangle$ correlator,
similar to the recent overlap results~\cite{Horvath:2005cv}.
Note that errors were calculated
using a first-order jackknife procedure but are too small to see.}
  \label{fig:qxqysweeps}
\end{figure}

Recent studies of vacuum structure in Lattice
QCD~\cite{deForcrand:2006my,Horvath:2005cv} have revolved around the
use of the overlap topological charge density
operator~\cite{Neuberger:1998my}. The overlap operator has the benefit
of producing an integer topological charge and was first to reveal the
negative topological charge density correlator, $\langle q(x)q(0)
\rangle$~\cite{Horvath:2005cv} for $x>0$.  Unfortunately, the overlap
operator is very computationally intensive.  Thus we examine the issue
of whether a traditional smearing method can produce a negative
correlator.  Using the quenched gauge fields, we calculate the
$\langle q(x)q(0) \rangle$ correlator on smoothed gauge fields with a
three-loop ${\mathcal O}(a^4)$-improved lattice field strength tensor
\cite{BilsonThompson:2002jk}.  
Fig.~\ref{fig:qxqysweeps} reports the results.

\subsection{Instanton-Like Objects}

Repeated application of  a smearing algorithm will eventually
reveal the presence
of spherical instantons in a gauge field. However, after only a small
number of sweeps, these objects tend to be
far from spherical. We now wish to investigate
the similarity of these objects to instantons.

Using over-improved smearing we analyze the action density of a smeared
field to determine the peaks of maximum action and fit the instanton
action density to our data. From this we can extract a size, $\rho$, for the
instanton.
We also extract the charge at the centre of the instanton-like
object, $q(x_0)$.  Thus, if there is good agreement between
the extracted $q(x_0)$ and that predicted by $\rho$, then we can say that
the object is locally representative of an instanton.

\begin{figure}
  \begin{center}
    \begin{tabular}{l r}
      \includegraphics[angle=90,width=0.45\textwidth]{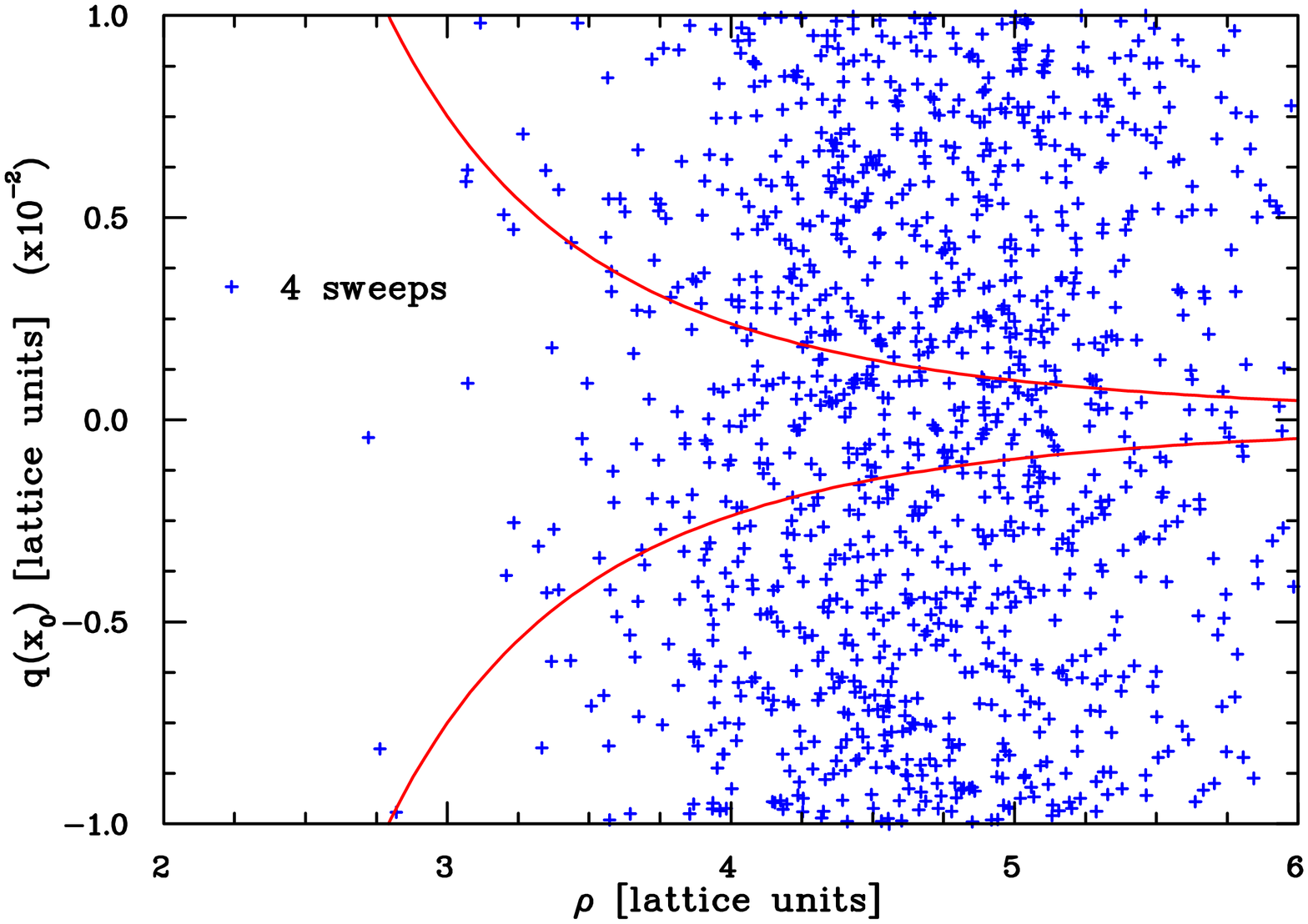}
      &
      \includegraphics[angle=90,width=0.45\textwidth]{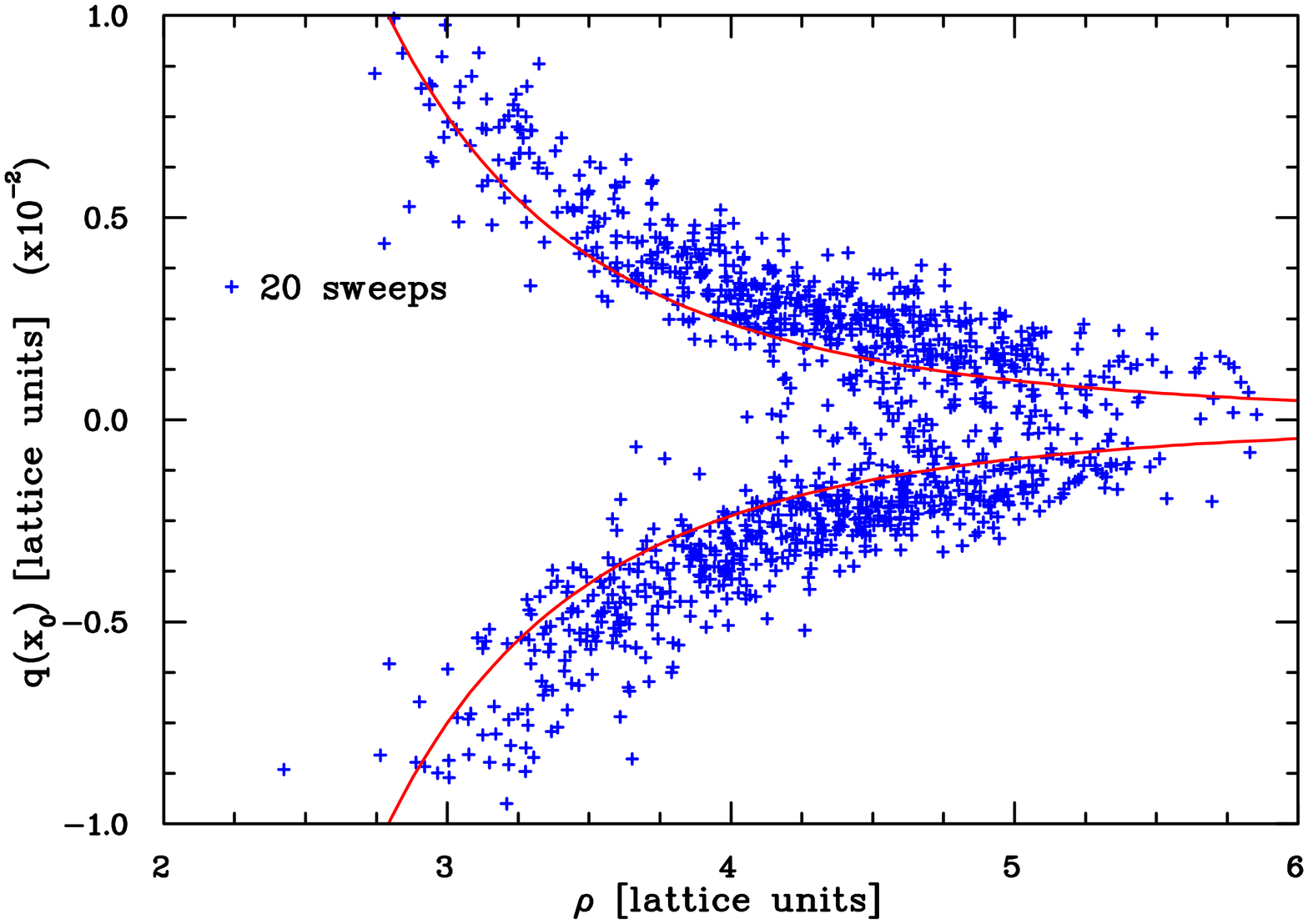}
    \end{tabular}
  \end{center}
  \caption{
    $q(x_0)$ versus the instanton size $\rho$ for 4 and 20 sweeps
    of over-improved stout-link smearing. Calculation details are in the text.
    We see that for 4 sweeps of
    smearing the peaks do not seem to represent instantons, but that as
    we smooth further the points start to cluster around the predicted line.
    \vspace{-12pt}
  }
  \label{fig:qvsrho}
\end{figure}

In Fig.~\ref{fig:qvsrho} we plot $q(x_0)$ vs $\rho$ for a quenched gauge field
after both 4 and 20 sweeps of over-improved stout-link smearing. 
Each cross represents a peak in the action density. If the peak were
to represent an instanton then its cross should lie on the theoretical
curve. We see that for 4 sweeps of smearing
the peaks do not appear to represent instantons. As more UV
fluctuations are removed the crosses lie closer to the line.

\subsection{The Vacuum Structure of Dynamical Gauge Fields}

Several studies of the differences in vacuum structure between
quenched and dynamical fields have focused on the topological
susceptibility
\begin{equation}
  \chi = \langle \int \! d^4 x \,\, q(x) \, q(0) \rangle = \frac{\langle Q^2 \rangle}{V}
\end{equation}
We now 
extend these studies of the topological susceptibility to the
$\langle q(x)q(0) \rangle$ correlator.
It has been shown that~\cite{Hart}
\begin{equation}
  \langle \int \! d^4 x \,\, q(x) \, q(0) \rangle \sim m_\pi^2
  \rightarrow 0 
  {\rm \ in \ the \ chiral \ limit,}
\end{equation}
and therefore $|Q| \rightarrow 0$ also. This leads to three scenarios
for how the shape of the $\langle q(x)q(0) \rangle$ could change in
the presence of dynamical quarks. Either the positive contact term and
the magnitude of the negative component could both increase or
decrease, or they could stay the same. The only requirement is that
the integral of the correlator vanishes in the chiral limit. However,
as we discussed in the introduction, fermion-loop coupling
renormalization leads to smaller $\beta$ admitting larger field
fluctuations.  We therefore expect that $\langle q(0)^2 \rangle$
should increase, and thus so must the negative component of $\langle
q(x)q(0) \rangle$ increase in magnitude.

\begin{figure}
  \begin{center}
    \includegraphics[angle=90,width=0.45\textwidth]{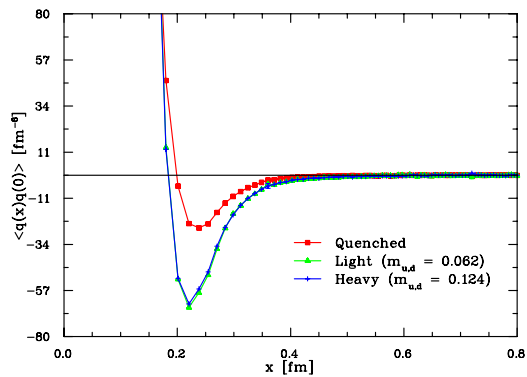}
  \end{center}
  \caption{The topological charge density correlator $\langle q(x)q(0)
\rangle$ for the quenched, light and heavy dynamical-fermion gauge
fields.  It is interesting to see how the dynamical fermion loops have
caused the negative dip to increase in magnitude, and how this effect
is greater for lighter quark masses.  Although not shown, the positive
contact term $\langle q^2(0) \rangle$ has also increased in
magnitude.  Exact values are given in the text.}
  \label{fig:qxqymultcfgs}
\end{figure}

Fig.~\ref{fig:qxqymultcfgs} shows the topological charge density correlator
as calculated for the three different types of gauge fields.
We see that the presence of dynamical quarks has caused the magnitude
of the negative component of the correlator to increase, and that this
effect is greater for lighter quark masses. The $x$-intercept has also
moved closer towards 0. Although not shown in the plot, the mean
square density $\langle q^2(0) \rangle$ has also increased in
magnitude.  The exact values of the positive contact term are;
quenched $= 2924\pm 4\, {\rm fm^{-8}}$, heavy $= 5251 \pm 12 \, {\rm
fm^{-8}}$, light $= 5432 \pm 8 \, {\rm fm^{-8}}$

\begin{figure}[t]
  \begin{center}
      \includegraphics[width=0.49\textwidth]{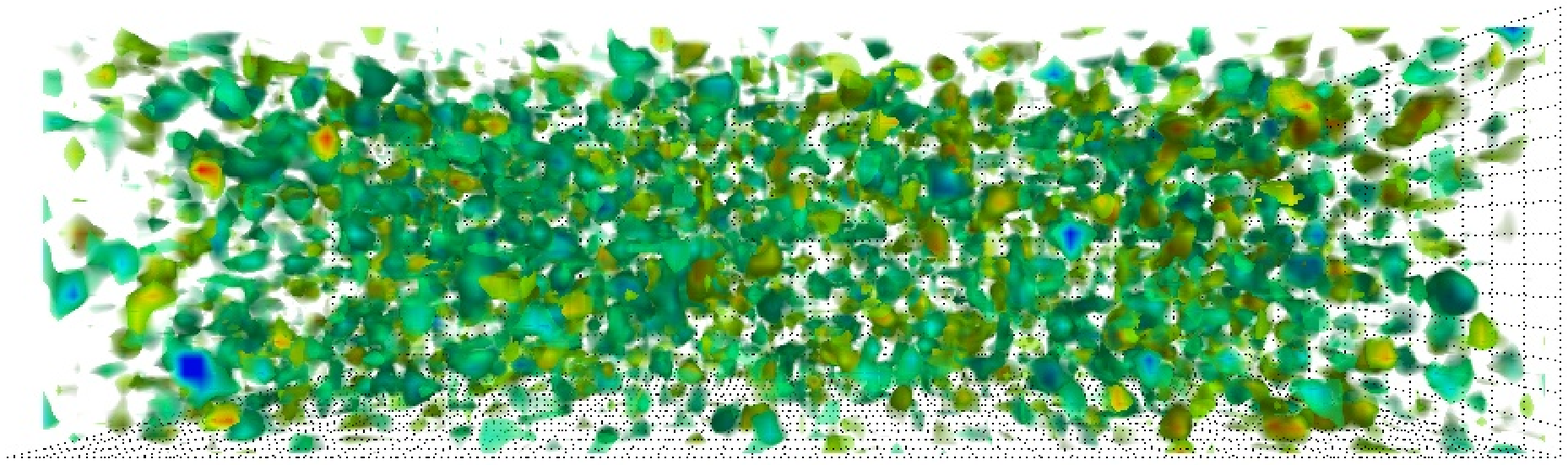} \\[-5pt]
      \includegraphics[width=0.49\textwidth]{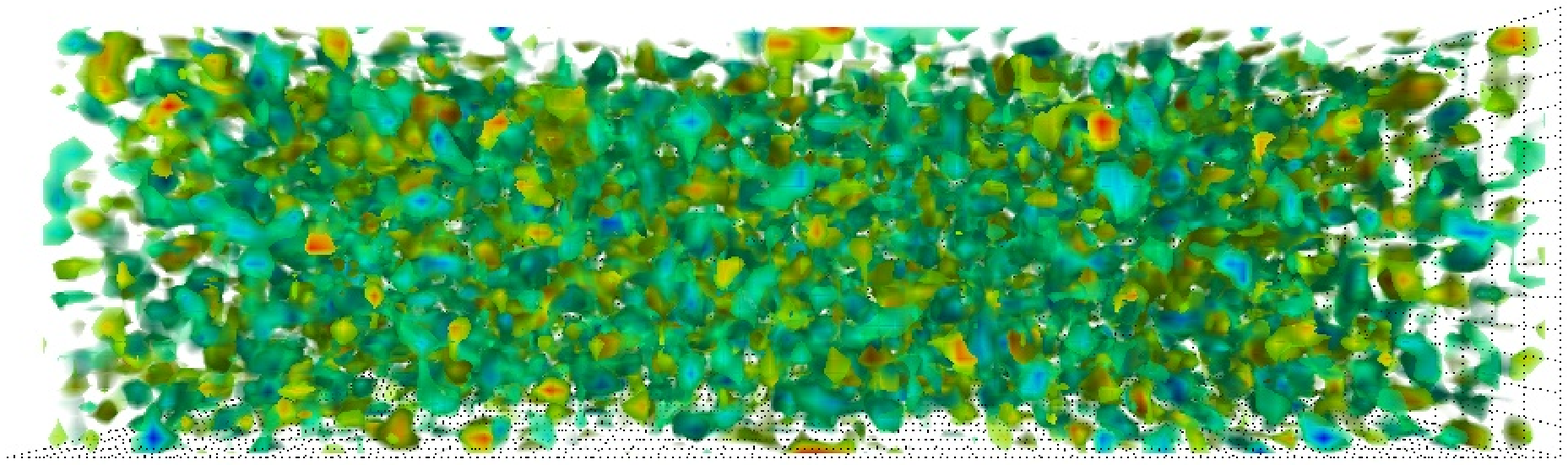}
      \includegraphics[width=0.49\textwidth]{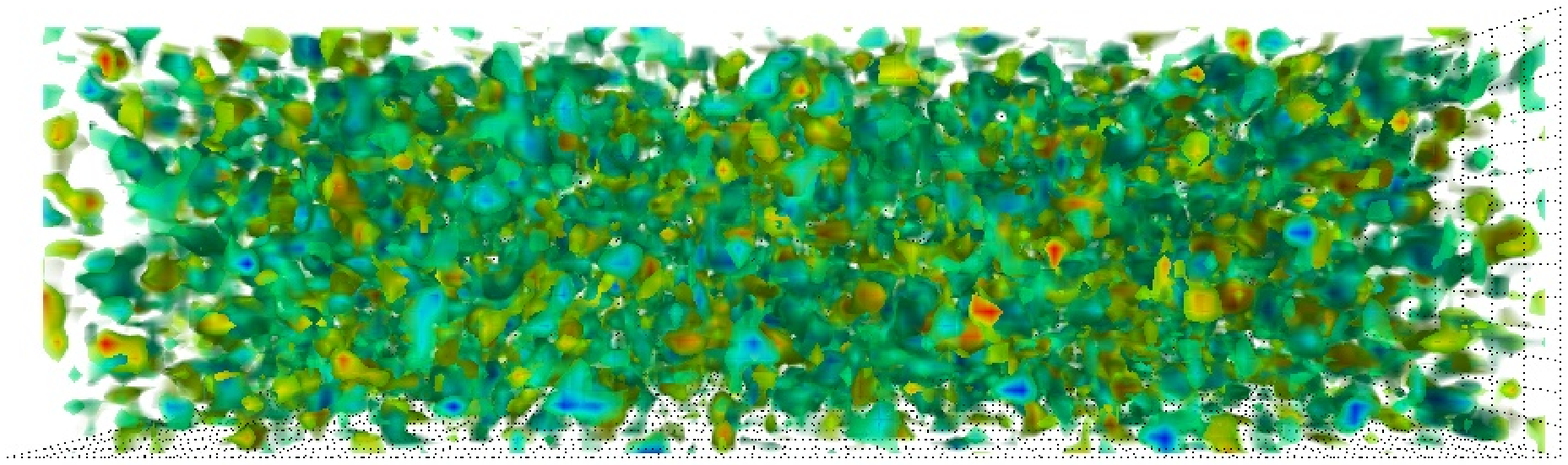}
  \end{center}
  \caption{ Topological charge densities for the quenched (top), heavy
    (bottom left) and light (bottom right) dynamical-fermion gauge
    fields.  Each field has been smoothed using 4 sweeps of
    over-improved stout-link smearing. We see that the bottom two
    dynamical gauge fields contain a higher density of non-trivial
    topological charge density than the quenched field.  However it is
    difficult to see the difference between the two dynamical gauge
    fields.  }
  \label{fig:chargedensities}
\end{figure}

We expect that this behaviour will be readily apparent
in visualizations of the charge density, $q(x)$. Plots of $q(x)$
are shown in Fig.~\ref{fig:chargedensities},
and indeed we observe that the light and heavy gauge fields contain many
more non-trivial field fluctuations. 

\section{Conclusion}

We have presented the first calculation of the topological charge
density correlator, $\langle q(x)q(0) \rangle$, in full QCD.  Using
both 3-loop improved cooling and the new over-improved stout-link
smearing procedure we are able to obtain negative $\langle q(x)q(0)
\rangle$ correlators. Using our proven methodology a quantitative
comparison of quenched and dynamical gauge fields is performed.  The
dynamical gauge fields show an increase in non-trivial vacuum field
fluctuations.  This is observed directly through visualizations of the
topological charge density and via the calculation of the $\langle
q(x)q(0) \rangle$ correlator. For the correlator we see an increase in
the magnitude of the negative dip and positive contact term with
larger effects for lighter quark masses.  These observations are in
accord with expectation, outlined in greater detail in a forthcoming
publication~\cite{MoranFuture}.


\begin{thebibliography}{99}

\bibitem{GarciaPerez:1993ki}
Margarita Garcia~Perez {\it et~al.},
\newblock {\em Nucl.\ Phys.} {\bf B413}, 535--552 (1994).
\newblock [{\tt hep-lat/9309009}]

\bibitem{Morningstar:2003gk}
Colin Morningstar and Mike~J. Peardon, 
\newblock {\em Phys.\ Rev.} {\bf D69}, 054501 (2004).
\newblock [{\tt hep-lat/0311018}]

\bibitem{Bonnet:2001rc}
Frederic D.~R. Bonnet {\it et~al.}, 
\newblock {\em Phys.\ Rev.} {\bf D65}, 114510 (2002).
\newblock [{\tt hep-lat/0106023}]

\bibitem{BilsonThompson:2002jk}
Sundance~O. Bilson-Thompson {\it et~al.},
\newblock {\em Ann.\ Phys.} {\bf 304}, 1--21 (2003).
\newblock [{\tt hep-lat/0203008}]

\bibitem{Bilson-Thompson:2001ca}
Sundance~O. Bilson-Thompson {\it et~al.},
\newblock {\em Nucl.\ Phys.\ Proc.\ Suppl.} {\bf 109A}, 116--120 (2002).
\newblock [{\tt hep-lat/0112034}]

\bibitem{Belavin:1975fg}
A.~A. Belavin {\it et~al.},
\newblock {\em Phys.\ Lett.} {\bf B59}, 85--87 (1975).

\bibitem{Bernard:2001av}
Claude~W. Bernard {\it et~al.},
\newblock {\em Phys.\ Rev.} {\bf D64}, 054506 (2001).
\newblock [{\tt hep-lat/0104002}]

\bibitem{Aubin:2004wf}
C.~Aubin {\it et~al.},
\newblock {\em Phys.\ Rev.} {\bf D70}, 094505 (2004).
\newblock [{\tt hep-lat/0402030}]

\bibitem{deForcrand:2006my}
Philippe de~Forcrand,
\newblock {\em AIP Conf.\ Proc.} {\bf 892}, 29--35 (2007).
\newblock [{\tt hep-lat/0611034}]

\bibitem{Horvath:2005cv}
I.~Horvath {\it et~al.},
\newblock {\em Phys.\ Lett.} {\bf B617}, 49--59 (2005).
\newblock [{\tt hep-lat/0504005}]

\bibitem{Neuberger:1998my}
Herbert Neuberger,
\newblock {\em Phys.\ Rev.\ Lett.} {\bf 81}, 4060--4062 (1998).
\newblock [{\tt hep-lat/9806025}]

\bibitem{Hart}
A.~Hart and M.~Teper, 
\newblock {\em Phys.\ Lett.} {\bf B523}, 280--292 (2001).
\newblock [{\tt hep-lat/0108006}]

\bibitem{MoranFuture}
Peter~J. Moran and Derek B. Leinweber,
\newblock {in preparation}, 2007.

\end{thebibliography}

\end{document}